# Room temperature two terminal tunnel magnetoresistance in lateral graphene transistor

C. I. L. de Araujo[1,2*], H. A. Teixeira[1], O. O. Toro[1], C. Liao[2], J. Borme[2], L. C. Benetti[2,3], D. Schafer[3], I. S. Brandt[3], R. Ferreira[2], P. Alpuim[2], P. P. Freitas[2], and A. A. Pasa[3*]

[1]Laboratory of Spintronics and Nanomagnetism (LabSpiN), Departamento de Física, Universidade Federal de Viçosa, Viçosa-MG, 36570-900, Brasil.

[2]International Iberian Nanotechnology Laboratory (INL), Ave. Mestre Jose Veiga, 4715-330 Braga, Portugal.

[3]Laboratório de Filmes Finos e Superfícies (LFFS), Universidade Federal de Santa Catarina, Florianópolis 88040-900, Brazil.

*Correspondence to: dearaujo@ufv.br, andre.pasa@ufsc.br

**Abstract:**

We investigate the behavior of both pure spin and spin-polarized currents measured with four probe non-local and two probe local configurations up to room temperature and under external gate voltage in a lateral graphene transistor, produced using a standard large-scale microfabrication process. The high spin diffusion length of pristine graphene in the channel, measured both directly and by the Hanle effect, and the tuning of relation between electrode resistance area present in the device architecture, allowed us to observe local tunnel magnetoresistance at room temperature, a new finding for this type of device. Results also indicate that while pure spin currents are less sensitive to temperature variations, spin-polarized current switching by external voltage is more efficient, due to a combination of the Rashba effect and change in carrier mobility by Fermi level shift

Spintronics devices have become very attractive for substitution of silicon-based technology since the huge improvements in magnetoresistive signal were obtained after the discovery of giant magnetoresistive effect (Baibich et al., 1988; Binasch et al., 1989) and magnetic tunneling at room temperature (Miyazaki eta Tezuka, 1995; Moodera et al., 1995). Magnetic tunnel junctions, at first based on amorphous insulators, reach nowadays very large tunnel magnetoresistive response, around 200 % at room temperature, by coherent tunneling through crystalline insulator (Butler et al., 2001; Parkin et al., 2004). Such high values enable the development of several families of magnetoresistive random access memories (Chappert et al., 2007) with nanoscale high density of bit storage (Hu et al., 2011) or multilevel storage in same bit cell (De Araujo et al., 2016). Those memories have been pointed out as potential substitutes for the current scalability limited capacitive devices, due to their non-volatile storage, speed, and protective characteristic against external doping (Åkerman, 2005). However, as the logic control of such stored medium is still realized by a metal oxide semiconductor field effect transistor, the scalability limitation in logic devices continues preventing further information technology improvements.

To surmount that limitation described above, the Spin Field Effect Transistor (SpinFET) architecture proposed by Data and Das (Datta eta Das, 1990), which would be formed by source and drain composed by ferromagnetic material injecting and extracting spin-polarized current in a semiconductor channel, could be used for integration of memory and logic spintronics devices. Semiconductors would continue to be used due to their long spin diffusion length in comparison with metallic materials. In the original proposal, the switching effect would be done by the gate voltage, allowing the spin mixing by the Rashba effect, which arises due to strong spin-orbit interaction in semiconductors (Bihlmayer et al., 2015). As the whole process would be governed by majority carriers, SpinFET technology is not limited by capacitive scalability and could present very fast operation. The first problem faced for the experimental realization of SpinFET devices with good sensitivity was the impedance mismatch between ferromagnetic materials and semiconductors (Schmidt et al., 2000), which cause large decay in the spin polarization through the interfaces with channel, source, and drain. Such a problem was originally circumvented by insertion of a tunneling barrier at the interfaces (Fert eta Jaffrès, 2001; Rashba, 2000). Despite several investigations on spin injection in semiconductors with non-local configurations, where two pairs of ferromagnetic/tunnel (FT) contacts are used, one for injecting the spin-polarized current promoting a spin concentration imbalance at the ferromagnetic/semiconductor interface, and the second one for measuring a voltage drop proportional to the pure spin current (Jansen, 2012), the insertion of the tunnel barrier and spurious magnetoresistive effects from the ferromagnetic electrodes have limited proper functioning of devices measured with local configuration in three-terminal devices, as originally suggested by Data and Das (Tombros et al., 2007). Those limitations are probably related to the narrow window among resistance areas (RA) of source ($R_SA$), drain ($R_DA$), and tunnel barrier ($R_BA$) that have to follow the relation $R_SA \leq R_BA \leq R_DA$ predicted by A. Fert and H. Jaffrès (Fert eta Jaffrès, 2001) and subsequently verified experimentally (De Araujo et al., 2013; Dlubak et al., 2012). Other difficulties related to local measurements in SpinFETs are the high spin-orbit interaction in semiconductors and depletion layers in source/drain-channel interfaces, which are responsible for high spin mixing rate (Appelbaum et al., 2007). After its first electrical characterization (Novoselov et al., 2016), graphene was rapidly pointed out as a promising material to avoid such limitations and be applied in the transistor channel due to its predicted very long spin diffusion length of about $\lambda_{SD}$ = 100 μm, weak spin-orbit coupling, and hyperfine interaction (Han et al., 2014; Pesin eta MacDonald, 2012; Yan et al., 2016). Recently a good number of works are stating that it is possible to measure SpinFETs with graphene channel at room temperature (Dankert eta Dash, 2017; Xia et al., 2010) and satisfactory spin polarization was recently measured at lower temperatures in a wide channel using local configuration measured with two contacts (Dlubak et al., 2012), such configuration is desirable to improve transistors density.

In this work, we have used a standard large-scale microfabrication process to fabricate a SpinFET based on buried ferromagnetic/tunnel electrodes and gate. This procedure allowed to measure pure spin and spin-polarized currents and to determine de spin diffusion length in the pristine graphene channel and the dependence of the magnetoresistance as a function of the gate voltage.

**Results and Discussions:**

For the fabrication of the SpinFET, we have developed an alternative architecture composed of buried permalloy/alumina FT contacts as source (S) and drain (D), with epitaxial graphene (pristine) grown by chemical vapor deposition technique on copper substrate and transferred to the device channel. The fabrication process started with a multilayer with $Al_2O_3$ 100/Ta 5/CuN 50/Ta 5/CuN 50/$Ni_{81}Fe_{19}$ 100 (thicknesses in nm) deposited by sputtering on a 200 mm p-doped silicon wafer. Contact geometries were defined by optical lithography and stacks were ion milled down to CuN layer with the assistance of mass spectrometer in-situ measurements. Successively, structures were again protected by photoresist, and the milling process down to the alumina bottom layer was processed to define the CuN gate line and electrical contacts. Following, a thicker layer of alumina was deposited all over the active device region and a planarization process was performed by ion milling up to ferromagnetic contacts being exposed. Finally, 2nm aluminum was deposited by sputtering and oxidized by oxygen plasma, to act as a tunnel barrier, before the graphene transference process being performed (Vieira et al., 2016). Such a configuration, presented in the cartoon of Figure 1a, was used to prevent any damage or impurity presence in the graphene layer due to the fabrication processes, once that graphene was transferred just in the last steps of the device processing. Devices were designed for local and non-local measurements with buried gate for transistor switching characteristics, as presented in Figures 1b, and also devices were fabricated with multiple contacts for measuring the properties as a function of the channel length. In the field emission scanning electron microscopy image of Figure 1c, the main region of the device is presented with the ferromagnetic tunnel (FT) contacts geometry and dimensions. The crosses represent regions where micro Raman spots were placed during graphene sheet characterization. Spectra obtained in those positions, presented in Figure 1d, show good quality of graphene sheet after transference. The tunneling characteristic of FT contacts can be observed in standard I x V curve characterization in Figure 1e and the symmetrical shape of the first derivative is presented in Figure 1f. The shape of the FT contacts and slight asymmetries of RA are engineered for an efficient spin-valve effect and compatibility with the resistance window match (Fert eta Jaffrès, 2001), necessary to allow efficient local configuration measurements. The circular shape characteristics of FT contacts were used for better results in the planarization step utilized in the buried contacts, such planarization is important to ensure the flatness of the graphene sheet after transference. Alumina was chosen as the tunnel barrier, due to its good compatibility with graphene (Dlubak et al., 2010) in comparison with crystalline MgO, which is also hydrophobic and then not so agreeable with our graphene transfer process.

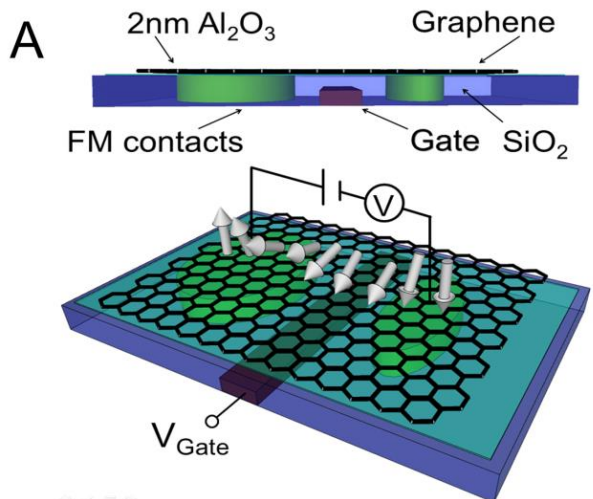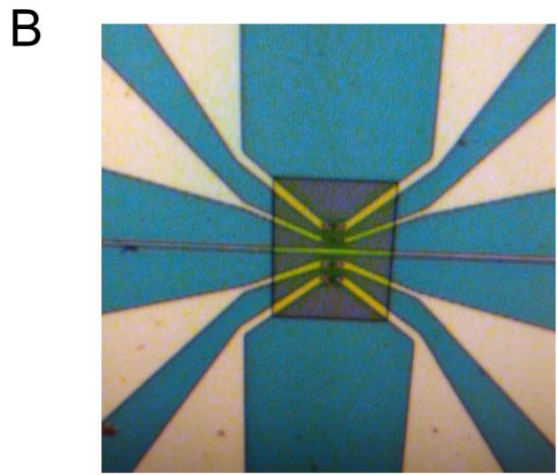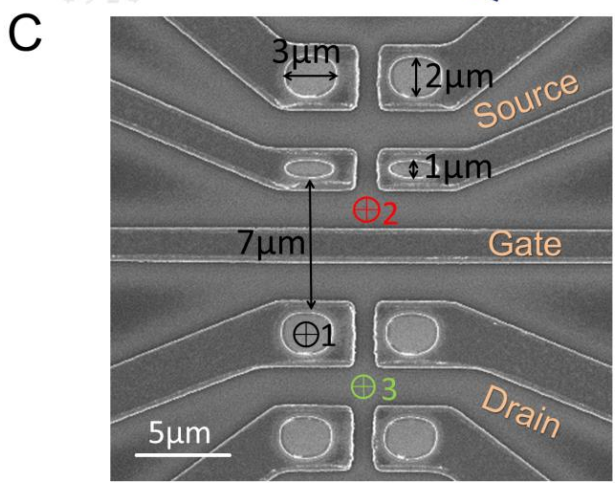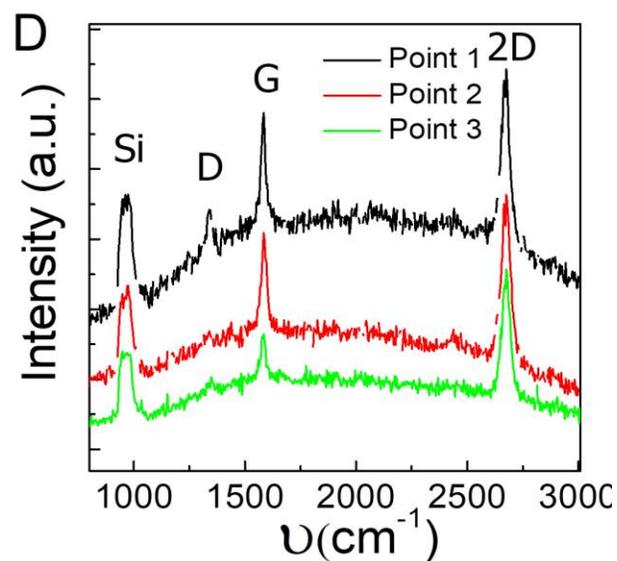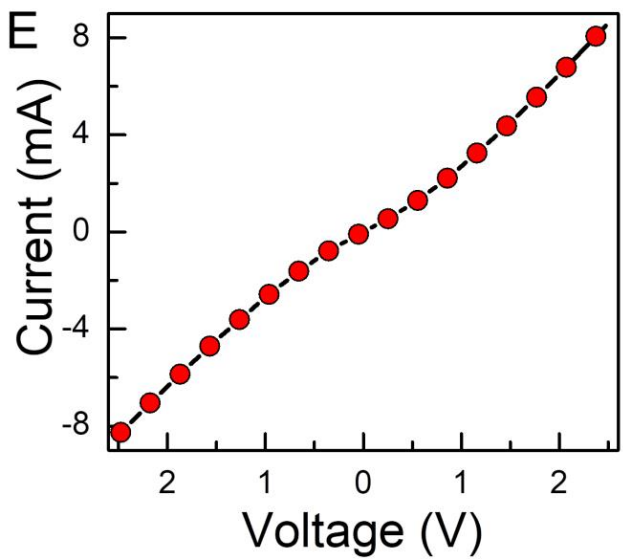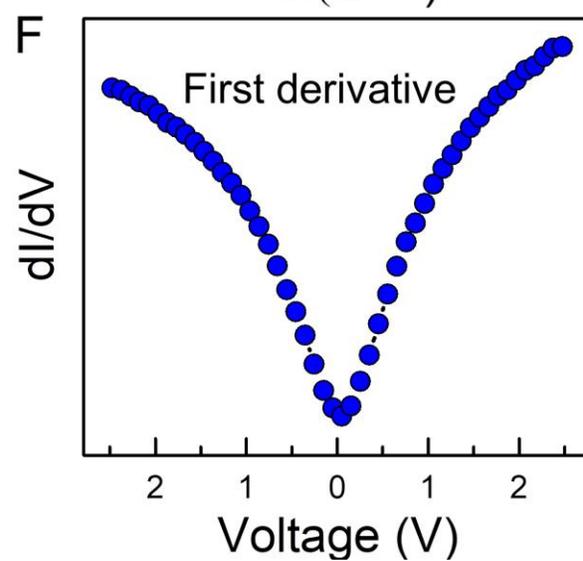

Figure 1 –a) Cartoon presenting the device architecture with ferromagnetic electrodes and gate buried in the silicon oxide, thin aluminum oxide for tunnel barrier, and epitaxial graphene sheet deposited on top, after milling sample planarization. b) transistor channel definition by plasma ashing after epitaxial graphene transference, c) field emission gun scanning electron microscopy image presenting a zoomed view of the buried ferromagnetic tunnel contacts Py/$Al_2O_3$, channel and gate with main dimensions and crosses representing the location where micro Raman spectroscopy was performed, d) Raman spectra from different sample location showing the good quality of graphene sheet and electrical characterization with e) I x V curve presenting standard tunnel behavior with f) reasonable symmetry in the first derivative.

After bonding with cold-pressed indium, transistors were inserted in a closed cycle helium cryostat for magnetoresistive characterizations under local (Figure 2a and 2b) and non-local configurations (Figure 2c and 2d) at room temperature. Design and geometries utilized here allowed pioneering the measurement of a good local magnetoresistive signal of 2.2 % at room temperature in a wide graphene channel of L = 7 μm. Using non-local configuration, the difference in resistance observed as a function of the external field was about 15 % at RT. Here it is important to discuss the unconventional TMR curve shape with non-abrupt resistance change and small hysteresis characteristics. Such behavior is proper of electrodes with disk shape of diameter d = 3 μm, which carries vortex magnetization in its ground state. In such disks, the magnetization reversal is not abrupt, as generally presented by ordinary rectangular electrodes, but smooth during magnetic vortex annihilation by the external field (De Araujo et al., 2014; Ribeiro et al., 2016). The next characterization performed was the measurement of local and non-local magnetoresistance signals under different temperatures, which allows the investigation of different scattering of pure spin and spin-polarized currents by phonons. In Figure 2e we present the decrease in the magnetoresistance as a function of temperature in the local configuration, while in non-local configuration the signal keeps constant with temperature. Such behavior can be understood by the fact that in local configuration there is a net spin-polarized charge and those charges are susceptible to scattering with generated phonons. Successive scatterings decrease the electron free path affecting the spin diffusion length in graphene. On the other hand, in the non-local configuration, the imbalance in the spin accumulation under the electrodes where the spin-polarized current is applied gives rise to a pure spin current that generates a spin accumulation and voltage drop dependent on the magnetic field in neighboring electrodes. That generated pure spin current between central FT contacts is much less susceptible to phonon scattering. The data presented in Figure 2e shows that at lower temperatures, where phonon scattering is negligible, both magnetoresistances measured for pure spin and spin-polarized currents have the same magnitude. That approximated constancy of pure spin current as a function of temperature is corroborated by recent works also performed in pristine graphene long channels (Chen et al., 2018; Kamalakar et al., 2015).

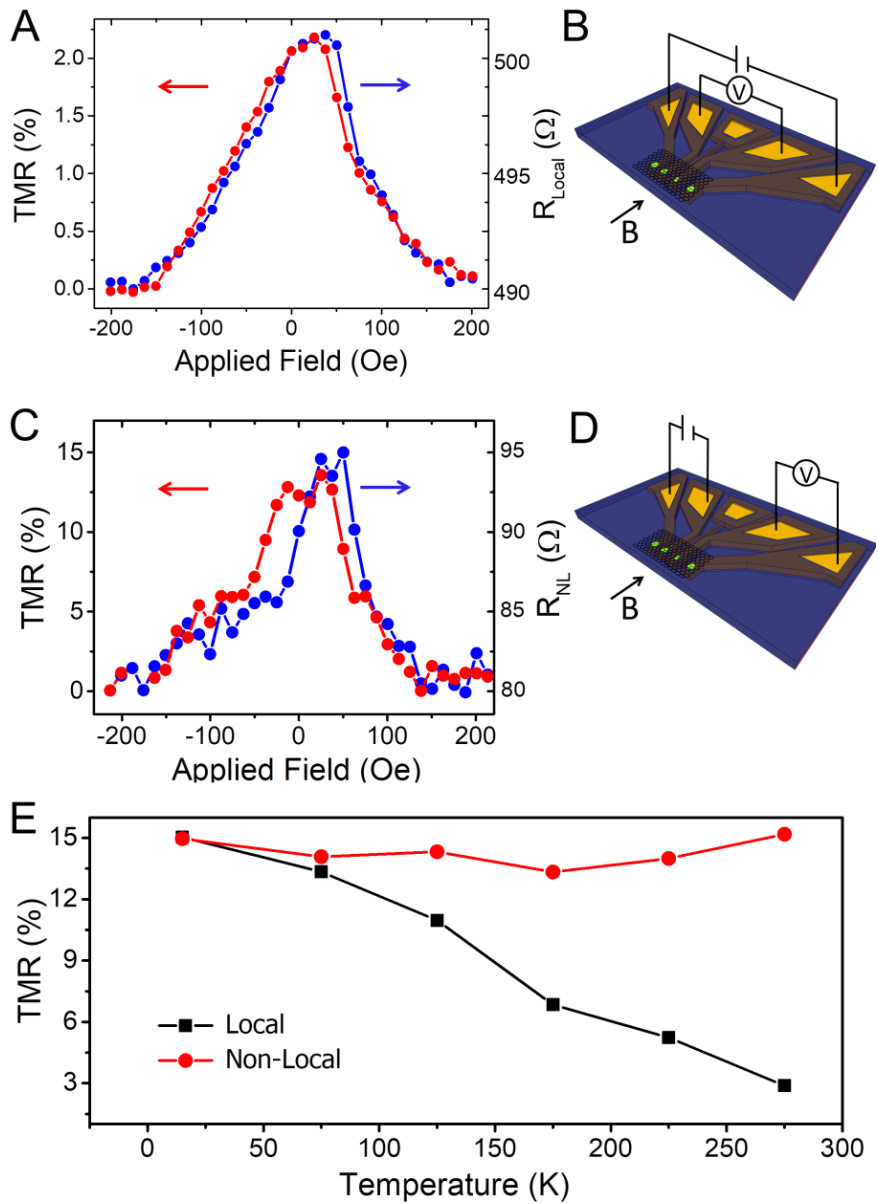

Figure 2 – **Tunnel magnetoresistance (TMR) and electrical resistance:** a) Room-temperature magnetoresistance local measurements as described in b) cartoon with four-point probe configuration, c) Non-local magnetoresistance measured following the configuration presented in d) with current applied in two contacts before gate and voltage drop measured in the next two contacts after the gate, as a function of the small external magnetic field. e) Presents the dependence of TMR in local and non-Local configuration as a function of temperature. The arrows indicate the direction of magnetic field scanning.

To characterize the spin lifetime in our devices at room temperature, we have performed Hanle effect measurements using a non-local configuration with an external field applied perpendicular to the graphene channel (Gurram et al., 2017). In the data presented in Figure 3a it is possible to observe that the amplitude is large and the curve is broad for positive applied current (spin extraction), in comparison with the negative curve (spin injection), such asymmetry is reported in the literature for biased Hanle measurements (Pi et al., 2010) and are related to the asymmetric spin accumulation due to the different RA in FT contacts. Such asymmetry is proportional to the one observed in the I x V curve presented in Figures 1e and 1f, which is also related to the different contact areas utilized to achieve spin-valve switching and fit the RA window, as mentioned above. As resistivity in the transferred graphene is characterized as ρ=3KΩ/□ and tunnel resistance as 500 Ω, these measurements are consistent with values presented in the literature for similar devices (Han et al., 2010; Han eta Kawakami, 2011), the resistance area window is accomplished by the geometries utilized in the present device as $R_SA = 5.4 K\Omega.\mu m^2 \geq R_SB = 3.0 K\Omega.\mu m^2 \geq R_SD = 0.8 K\Omega.\mu m^2$.

By fitting the data in Figure 3a with equation 1 presented below, from references (Han et al., 2010; Pi et al., 2010), relaxation times in the range of τ = 0.5 - 1.1 ns and diffusion coefficients of D = 0.0103 - 0.0142 $m^2s^{-1}$ are estimated.

$$\Delta V_{NL} \propto \pm \int_0^\infty \frac{1}{\sqrt{4\pi Dt}} \exp\left[\frac{-L^2}{4Dt}\right] \cos(w_L t) \exp\left(\frac{-t}{\tau}\right) dt, \qquad (1)$$

where $V_{NL}$ is the non-local measured voltage, the sign + (-) is for magnetization parallel (antiparallel), and $\omega_L = (g\mu_B H_\perp)/\hbar$ is the Larmor frequency and $H_\perp$ is the applied out-of-plane magnetic field. High carrier mobility $\mu_{FE}$ = 5200-7800 cm$^2$V$^{-1}$s$^{-1}$ of pristine graphene applied to Einstein relation, allows the calculation of charge diffusion coefficients around $D_C$ = 0.13 - 0.2 m$^2$s$^{-1}$. The estimated spin diffusion coefficient from Hanle measurements is one order of magnitude lower than the charge diffusion length calculated above, similar behavior is also observed in literature (Kamalakar et al., 2015). Using the equation for spin diffusion length, $\lambda_{SD} = \sqrt{D\tau}$, values in the range of $\lambda_{SD}$ = 2 – 4 µm can be estimated for the channel length of L = 7 µm. Now, in Figure 3b we present a direct characterization of spin lifetime, by fitting the experimental data of TMR decay measured in local configuration as a function of channel length, with the spin diffusion equation TMR ≈ exp(L/$\lambda_{SD}$) (Józsa et al., 2008; Valet eta Fert, 1993). Using this approach, we have directly measured a spin diffusion length of $\lambda_{SD}$ = 6.16 µm at room temperature in our pristine graphene channel, in good agreement with the value presented above indirectly estimated from the Hanle effect measurements.

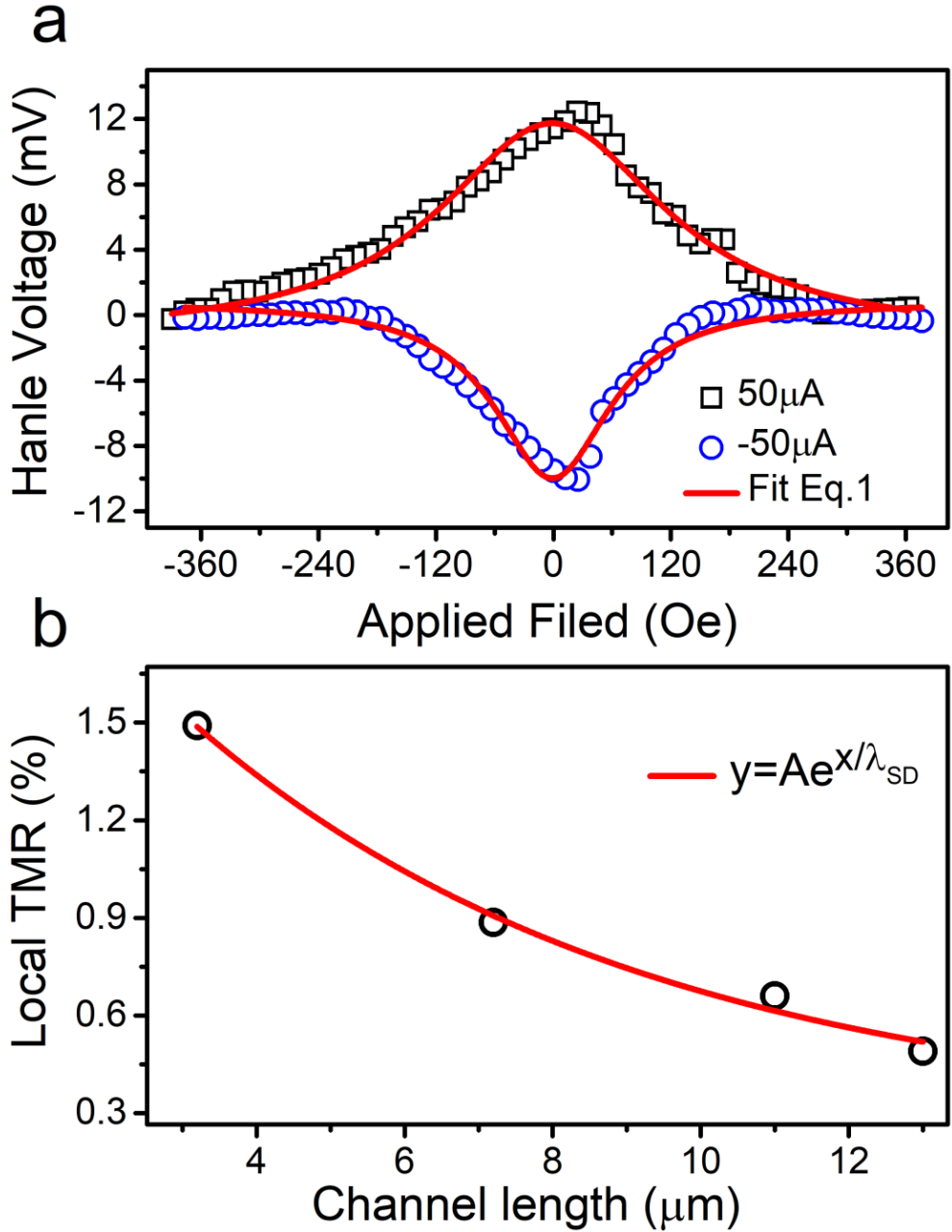

Figure 3 – Spin relaxation time and diffusion length: a) Hanle measurements performed in non-local configuration with spin injection in black squares (positive current) and detection in blue circles (negative current) and b) Local TMR decay as a function of channel length L.

To investigate the different response of pure spin current and spin-polarized current to the electric field in our graphene channel lateral device, we have performed measurements of non-local and local TMR signal as a function of gate voltage at the low temperature of 15 K. The evolution of non-local TMR curves at this temperature is presented in Figure 4a. The signal decay as a function of gate voltage is summarized in Figure 4b. We have also investigated, in an unprecedented way, the local TMR signal as a function of gate voltage. Curves of TMR measured at 0, 0.05, and 0.50 V are displayed in Figure 4c and a sharp signal decays as a function of gate voltage is presented in Figure 4d. Comparing both measurements configuration it is clear that gate voltage is much more effective in spin mixing of spin-polarized current, measured by local configuration, in comparison with non-local pure spin current measurement. In the original prediction by Data and Das, the main idea for spin switching was based on the Rashba effect, which would arise from spin-orbit interaction in the channel. However, in graphene, the spin-orbit interaction is predicted to be very weak and one could expect that spin mixing in our SpinFET could mainly occur due to an increase in hole/electron carrier concentration due to Fermi level shift under applied gate voltage. Previous results show that pure spin currents measured in non-local configuration, in long channel pristine graphene with sparse rippling, are not so sensitive to such carrier concentration increase (Kamalakar et al., 2015). The decrease in pure spin current relaxation time measured in epitaxial graphene, under external gate voltage, is mainly attributed to the spin-orbit effect imposed by hybridization of the $p_z$ orbitals with $p_x$ and $p_y$ orbitals from the σ-band (Huertas-Hernando et al., 2006), due to sheet ripples after transfer to substrate (Han et al., 2014) or proximity effect with capping layer as $MoS_2$ (Dankert eta Dash, 2017). In the present work, the decay of pure spin current as a function of external gate voltage presented in Figure 4b is very smooth in comparison with our measurement as a function of gate voltage in three-terminal devices shown in Figure 4d. That effective signal decay in such a configuration could be explained by a combination of Rashba effect, owing to the graphene ripples, with a decrease in electron mean free path and consequently spin mixing, due to an increase of carrier concentration with Fermi level shift under external gate voltage. This sharp decrease in local magnetoresistive signal is desirable for fast and low power consumption devices.

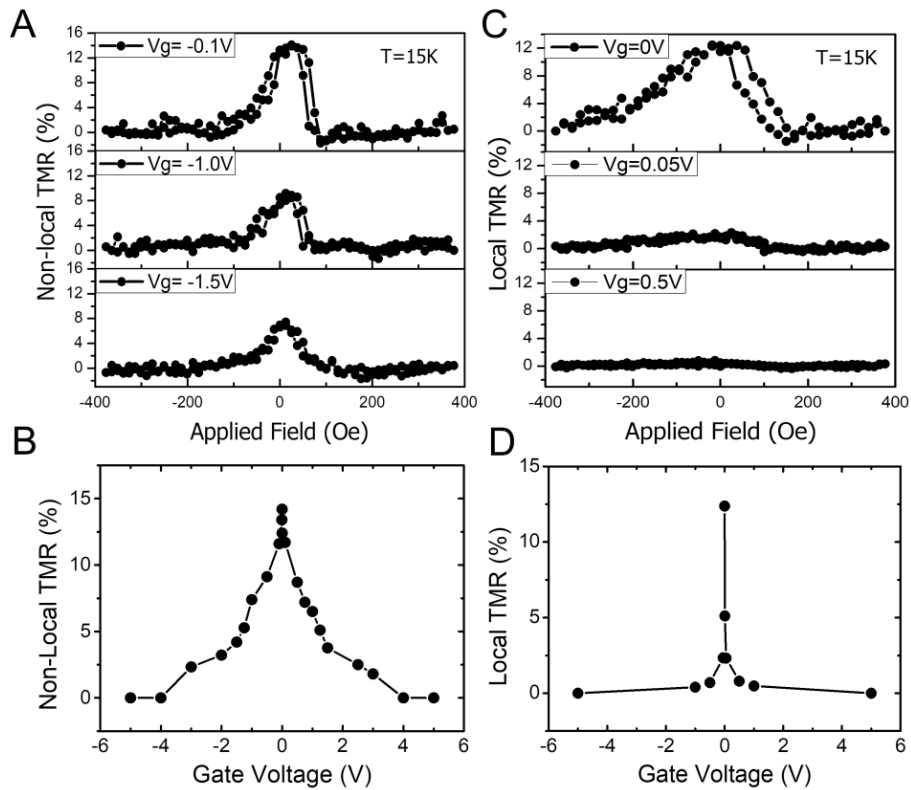

Figure 4 – **TMR measurements as T = 15K as a function of magnetic field and gate voltage:** a) Non-local magnetoresistance curves measured as a function of the magnetic field, b) non-local TMR slow drop as a function of gate voltage, c) local magnetoresistance curves measured as a function of the magnetic field, and d) local TMR fast drops as a function of gate voltage.

**Conclusions:**

In conclusion, we have successively applied a standard large-scale microfabrication process to fabricate a SpinFET based on buried ferromagnetic/tunnel electrodes and gate. The fabricated devices with electrodes engineered to fit the relation between tunnel barrier and channel resistance areas presented considerable pure spin and spin-polarized current injection and extraction at low and room temperature, which could be used to estimate directly and indirectly the spin diffusion length in the utilized pristine graphene channel at room temperature. The particular response of pure spin current and spin-polarized current to temperature, spin-orbit coupling imposed by sheet ripples, and carrier concentration modification by Fermi level shift through gate voltage, could be investigated and related with net charge sensitivity to phonons and electrical carrier scatterings. The results presented here suggest the feasibility of large-scale three terminals SpinFET and its application as a future logic device operating at room temperature.

**Acknowledgments:**

The financial support for this research was provided by the Brazilian agencies FINEP, FAPEMIG, and CAPES (Finance Code 001).